\documentclass[amsmath, amssymb, amsfonts, twocolumn, prl, 10pt]{revtex4-1}
\usepackage{bbold}
\usepackage{tikz, adjustbox}
\usepackage{graphicx}

\usepackage{microtype}
\newcommand{\dd}{\mathrm{d}}

\begin{document}

\title{Carroll Geometry Meets De Sitter Space via Holography}

\author{Chris D. A. Blair${}^1$, Niels A. Obers${}^{2, 3}$, and Ziqi Yan${}^{2, 3}$ \smallskip\smallskip}

\affiliation{\smallskip%
${}^1$Instituto de F\'{i}sica Te\'{o}rica UAM/CSIC, Universidad Aut\'{o}noma de Madrid, Cantoblanco, Madrid 28049, Spain \smallskip\\
${}^2$Center of Gravity, The Niels Bohr Institute, Copenhagen University, Blegdamsvej 17,
DK-2100 Copenhagen \O, Denmark \smallskip\\
${}^3$Nordita, KTH Royal Institute of Technology and Stockholm University,
Hannes Alfv\'{e}ns v\"{a}g 12, SE-106 91 Stockholm, Sweden
}

\begin{abstract}
We explain how to relate the ideas of Carroll geometry, matrix theory on instantonic objects, and infinite boost limits of M-theory. Based on these new insights, we explore the implications for possible holographic constructions involving a de Sitter or flat space bulk. We show that Carroll-like (including particle Carroll) geometry in a hypothetical de Sitter holography mirrors the recently realized important role played by Galilei-like geometry in matrix theory and the AdS/CFT correspondence. This also allows us to generate examples of holography with a Carroll-like bulk.      
\end{abstract}

\maketitle

\begin{figure*}[t!]
    \centering
    \includegraphics[scale=0.9]{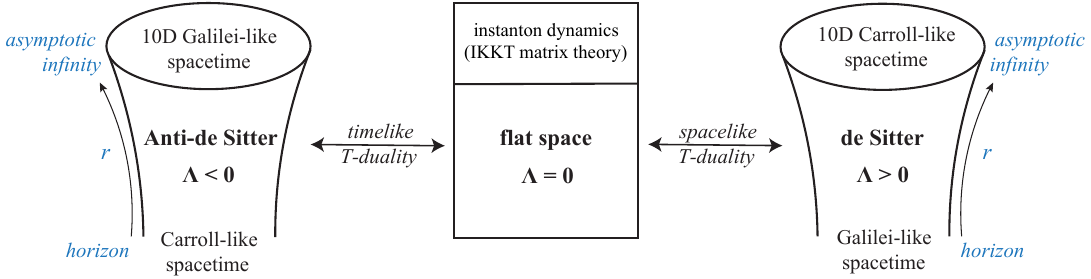}
    \caption{Timelike T-duality as a looking-glass: it maps between not only Galilei- and Carroll-like geometries but also AdS and dS holography. The AdS (dS) bulk geometry itself also acts as a looking-glass between Galilei-like (Carroll-like) geometry at the asymptotic infinity and Carroll-like (Galilei-like) geometry on the horizon.}
    \label{fig:rm}
\end{figure*}


A remarkable milestone in modern physics was the discovery that our Universe is currently undergoing accelerated expansion~\cite{SupernovaSearchTeam:1998fmf, SupernovaCosmologyProject:1998vns}.
This implies the existence of a small and positive cosmological constant, and hence that the Universe will asymptotically approach a de Sitter (dS) space.
Associated with these implications are prominent puzzles. 
The observed cosmological constant is hierarchically much smaller than any prediction from effective field theory. 
It is therefore well motivated to understand how dS space could arise from a self-consistent quantum gravity, such as string theory. However, this is notoriously difficult; in fact, there is also evidence indicating that dS space may be quantum mechanically unstable (see \emph{e.g.}~\cite{Maldacena:2000mw, Goheer:2002vf, Dvali:2014gua, Danielsson:2018ztv, Brennan:2017rbf}). This instability of dS space may be further supported by the recent hints that dark energy evolves over cosmic time~\cite{DESI:2025zgx, DES:2025bxy}. Nevertheless, the area entropy law associated with the dS cosmological horizon~\cite{Spradlin:2001pw} still suggests there could be a holographic description, relating the physics of dS space to a lower-dimensional dual field theory.
Such a correspondence would mirror the celebrated framework of holography for anti-de Sitter (AdS) spaces, with negative cosmological constant, which in particular is supported by explicit string theory realisations \cite{Maldacena:1997re}.

At small scales, our Universe can be approximated by Minkowski space with zero cosmological constant. 
This setting has also seen a search for a holographic description.
For 4D Minkowski space, this has stimulated the active program of flat space holography. There are two parallel approaches~\cite{Pasterski:2021raf, Donnay:2022aba, Bagchi:2022emh}: in \emph{celestial} holography, the dual conformal field theory (CFT) is proposed to live on the celestial sphere; in \emph{Carrollian} holography, the CFT lives on the (2+1)D null infinity and enjoys a Carrollian symmetry arising from a zero speed of light limit of the Poincar\'{e} group. In addition, examples using twistor space also appear to be useful~\cite{Costello:2022wso}. Even though copious progress has been made on the gravity side, the precise formulation of the postulated dual CFT still remains mysterious.

It is perhaps not fully appreciated that string theory also offers perspectives on flat space holography.
It was conjectured by Banks, Fischler, Shenker, and Susskind (BFSS)~\cite{Banks:1996vh} that M-theory in flat 11D spacetime corresponds to the large $N$ limit of a quantum mechanical system of $N \times N$ matrices. BFSS matrix theory arises from a BPS decoupling limit of a stack of $N$ D0-branes that are particle states in type IIA superstring theory~\footnote{From a IIA perspective, BFSS model at large $N$ also describes a (conformally) AdS${}_2$ dual geometry, which becomes invalid in the deep bulk where the M-theory description is instead required.}.
In this limit, the mass of a background D0-brane is blown up, while its charge is fine tuned to cancel any divergences in all physical observables. 
Historically, a direct link between BFSS matrix theory and 11D M-theory follows by performing a matrix regularization of the supermembrane~\cite{deWit:1988wri}. An analogous regularization of the IIB Green-Schwarz superstring leads to Ishibashi-Kawai-Kitazawa-Tsuchiya (IKKT) matrix theory~\cite{Ishibashi:1996xs}, which is in a way related to BFSS matrix theory via a timelike T-duality transformation~\footnote{The original IKKT matrix theory is derived from matrix-regularising the type IIB superstring and provides a non-perturbative definition of the IIB theory. Strictly speaking, this is distinct from the the IKKT matrix theory obtained from performing a timelike T-duality transformation in BFSS matrix theory, which leads us to type IIB${}^*$ string theory with a distinct Ramond-Ramond sector. This paper focuses on the latter case in the IIB${}^*$ theory.}. 
The holographic dual geometry of IKKT matrix theory should be 10D flat spacetime~\cite{Bergshoeff:1998ry, Ooguri:1998pf}. 
Such glimpses of flat space holography from matrix models have in part motivated a recent revival of interest in these theories, \emph{e.g.} \cite{Miller:2022fvc, Tropper:2023fjr, Herderschee:2023bnc, Komatsu:2024vnb, Hartnoll:2024csr, Komatsu:2024bop, Komatsu:2024ydh, Ciceri:2025maa}.

In this Letter, we take seriously these hints from string theory and propose a series of relations between (A)dS and flat space holography, Carroll geometry, and matrix theory.
We build on recent progress in understanding the nature of BPS decoupling limits in string theory \cite{Blair:2023noj, Gomis:2023eav} and related holographic constructions \cite{Lambert:2024uue, Lambert:2024yjk, Fontanella:2024kyl, Blair:2024aqz, Lambert:2024ncn, Harmark:2025ikv, Guijosa:2025mwh, Blair:2025prd}.
These limits lead to non-Lorentzian backgrounds, \emph{i.e.}~geometries with \emph{no} 10D metric.
This results in the appearance of an asymptotic Galilei-like geometry in AdS/CFT~\cite{Blair:2024aqz}, which is ultimately tied to the BPS decoupling limit that gives BFSS matrix theory. Holographically, such a limit corresponds to the near-horizon limit in the bulk that leads to \emph{e.g.}~the AdS geometry, and the non-Lorentzian geometry is realised in the asymptotic infinity of this near-horizon geometry~\cite{Blair:2024aqz}.

We will extend this picture to hypothetical dS holography in string theory (see Fig.~\ref{fig:rm}).
Elaborating on older observations of Hull~\cite{Hull:1998vg}, the key point is that the flat space correspondence of IKKT matrix theory is T-dual to a 10D geometry containing a conformally dS space as a subspace. Gravity on this geometry is unstable, and its field theory dual lives on instantonic objects. Such instability is a generic feature of timelike T-duality of the standard AdS/CFT correspondence. 
We will show that, in this hypothetical dS holography, the geometry at the asymptotic infinity is Carroll-like.

In M-theory the BPS decoupling limits discussed above are related to a particular infinite boost limit~\cite{Banks:1996vh}. We will show that there is an opposite infinite boost limit that zooms in on the center of the bulk geometry and leads to a `non-BPS' version of non-Lorentzian geometry. 

\vspace{0.5em}\noindent\textbf{From de Sitter to Carroll.}
We start with a close analog of the AdS${}_5 \times S^5$ geometry in the AdS${}_5$/CFT${}_4$ correspondence, where AdS${}_5$ is replaced with dS${}_5$ and the sphere $S^5$ is replaced with the hyperbolic space $H^5$~\cite{Hull:1998vg}. This $\text{dS}_5 \times H^5$ geometry is described by the metric
\begin{align} \label{eq:dsto}
	\dd s^2 &= \frac{ - \dd x_0^2 + \dd x_{5}^2 + \cdots + \dd x_9^2}{\mathbb{\Omega}}
     		+ \mathbb{\Omega} \, ( \dd x_1^2 + \cdots + \dd x_4^2 )\,, \notag \\[4pt]
     	\mathbb{\Omega} & = \left( \frac{\tau}{\ell} \right)^{\!2}\!,\quad \tau^2 = x_0^2 - r^2    ,
     		\quad%
	r^2 = x_5^2 + \dots + x_9^2\,.
\end{align}
We focus on the branch with $\tau^2 > 0$\,. To verify that this metric is manifestly $\text{dS}_5 \times H^5$\,, we perform the change of variables
$x_0 = \ell \, e^{t} \, \cosh \beta$ and
$r = \ell \, e^{t} \, \sinh \beta$\,,
under which $\tau = \ell \, e^t$ and the metric becomes
\begin{align}
\begin{split}
	\dd s^2 = & - \ell^2 \, \dd t^2 + e^{2 \, t} \, \bigl( \dd x_1^2 + \cdots + \dd x_4^2 \bigr) \qquad \textbf{dS${}_\mathbf{5}$} \\[4pt]
	& + \ell^2 \, \bigl( \dd \beta^2 + \sinh^2 \! \beta \, \dd \Omega_4^2 \bigr)\,\, \qquad\qquad\quad\, \mathbf{H^5}
\end{split}
\end{align}
The above geometry, supplemented with the Ramond-Ramond (RR) field $C^{(4)}$ and dilaton field $\Phi$ with
\begin{align} \label{eq:mmftrrgs}
	C^{(4)} = \mathbb{\Omega}^2 \, g^{-1}_\text{s} \, \dd x^1 \cdots \dd x^4\,,
		\quad%
	e^\Phi = g^{}_\text{s}\,,
        \quad%
    \ell^4 \sim N g^{}_\text{s}\,,
\end{align}
initially arises as a solution to the IIB${}^*$ supergravity, which is related to IIA supergravity via timelike T-duality. A characteristic feature of II${}^*$ supergravity is that the RR kinetic terms acquire a `wrong' sign, indicating instabilities. The `boundary' field theory analogous to $\mathcal{N} = 4$ super Yang-Mills theory (SYM) in AdS${}_5$/CFT${}_4$ is now a Euclidean SYM theory on a stack of E(uclidean)-branes~\footnote{E-branes are sometimes also referred to as S(pacelike)-branes~\cite{Gutperle:2002ai}.} extending in $x^i$, $i = 1,\cdots,4$. Denote the transverse coordinates as $x^a$, $a =0,5,\cdots,9$, we write in static gauge
\begin{align} \label{eq:efba}
\begin{split}
	S_\text{E4} = T_4 \int \dd^4 \sigma  & \, \text{tr} \Bigl( \tfrac{1}{2} \, D_i X^a \, D_i X_a + \tfrac{1}{4} \, F_{ij} \, F^{ij} \\[4pt]
		& - \tfrac{1}{4} \, [X^a,X^b] \, [X_a,X_b] \Bigr) + \text{fermions},
\end{split}
\end{align}
with $\sigma^i$ the worldvolume coordinates and $X^a$ the matrix version of the embedding coordinates transverse to the E4-branes. 
This field theory dual contains ghosts and may be non-unitary~\footnote{It has been suggested that the theory may be made well defined by taking into account massive higher spin states~\cite{Hull:1998vg} or using supergroups~\cite{Dijkgraaf:2016lym}.}. This pathology should be viewed as a feature of a dS holography: a bulk correlation function between operators residing on the future (past) infinity of the proper time in general corresponds to boundary conformal blocks with complex conformal weights~\cite{Strominger:2001pn, Balasubramanian:2002zh}.    

In the regime far away from the light-cone $x_0^2 = r^2$, we have $\mathbb{\Omega} \rightarrow \infty$ and the geometry in Eq.~\eqref{eq:dsto} becomes Carroll-like. One simple way to illustrate the Carroll-like boost symmetry is by using a fundamental (F1-)string to probe this asymptotic geometry. Denote the worldsheet coordinates as $\sigma^\alpha = (\tau,\sigma)$ and embedding coordinates $(X^a, X^i)$. In conformal gauge, the bosonic part of the Polyakov action can be found following~\cite{Gomis:2023eav} to be
\begin{align} \label{eq:fsa}
    S = \! \frac{T}{2} \! \int \! \dd^2 \sigma \Bigl( \partial_\tau X^a \, \partial_\tau X_a - \partial_\sigma X^i \, \partial_\sigma X^i + \lambda_i \, \partial_\tau X^i \Bigr).
\end{align}
See End Matter for a derivation of this action~\footnote{For related Carrollian actions, see~\cite{Cardona:2016ytk, Bagchi:2023cfp, Bagchi:2024rje, Harksen:2024bnh}.}.
This worldsheet theory is invariant under the target space Carroll-like boost
$\delta_\text{C} X^a = \beta^a{}_i \, X^i$ and
$\delta_\text{C} X^i = 0$\,,
with the spatial sector $X^i$ absolute. 
The worldsheet geometry is also Carrollian, which, in imaginary time, is topologically equivalent to nodal Riemann spheres~\cite{Gomis:2023eav}. 

Therefore, the E4-branes described by the action \eqref{eq:efba} live in a 10D Carroll-like spacetime that arises asymptotically, with the brane worldvolume directions forming an absolute spatial sector. 
Borrowing intuition from AdS/CFT, the `string picture' with weak 't Hooft coupling is now analogously described by a limit zooming in on a background E4-brane within type IIB${}^*$ superstring theory in flat spacetime. This limit is prescribed by reparametrising the background fields using a constant $\omega$ as (\emph{i.e.}~replace $\mathbb{\Omega}$ with $\omega$ in Eqs.~\eqref{eq:dsto} and \eqref{eq:mmftrrgs})
\begin{align} \label{eq:mm5tl}
\begin{split}
	\dd s^2 &= \tfrac{1}{\omega} \, \dd x^a \, \dd x^{}_a + \omega \, \dd x^i \, \dd x^i, 
        \quad%
    i = 1,\cdots, 4, \\[4pt]
     C^{(4)} &= \tfrac{\omega^2}{g^{}_\text{s}} \, \dd x^1 \cdots \dd x^4\,,
		\qquad\quad\,\,\,\,%
	e^\Phi = g^{}_\text{s}\,,
\end{split}
\end{align}
while the $B$-field and other RR potentials are set to zero. 
This $\omega \rightarrow \infty$ limit of the IIB${}^*$ theory leads to \emph{Matrix E4-brane Theory} (ME4T)~\footnote{ME$q$T is referred to as Matrix $p$-brane Theory (M$p$T) with $p = - q - 1$ in ~\cite{Gomis:2023eav, Blair:2023noj}.}, in which the fundamental degrees of freedom are captured by the E4-branes in 10D Carroll-like spacetime~\cite{Gomis:2023eav, Blair:2023noj}. Notably, the infinite $\omega$ limit of type II${}^*$ string theory is not only applied to the geometric data but also the dynamics.

This limit in the asymptotic regime can be phrased as a rescaling $\hat{x}^a = x^a/\sqrt{\omega}$ and $\hat{x}^i = \sqrt{\omega} \, x^i$, which also applies to the bulk back-reacted E4-brane geometry, 
\begin{align}
\label{eq:efbg}
    \dd \hat{s}^2 &= \hat{H}^{\frac{1}{2}} \, \dd \hat{x}^a \, \dd \hat{x}^{}_a + \frac{\dd \hat{x}^i \, \dd \hat{x}^i}{\hat{H}^{\frac{1}{2}}}\,, 
        &%
    \hat{H} &= 1 + \frac{\ell^4}{(\hat{x}^a \, \hat{x}_a)^2}\,,\notag \\[4pt]
    \hat{C}^{(4)} &= g^{-1}_\text{s} \, \hat{H}^{-1} \, \dd \hat{x}^1 \cdots \dd \hat{x}^4\,,
        &%
    e^{\hat{\Phi}} &= g^{}_\text{s}\,,
\end{align}
with $H$ the harmonic function and  $\ell^4 \sim N \, g^{}_\text{s}$ constant, assuming that $N$ is large so the geometric description is valid.
Plugging in the reparametrization that gives~\eqref{eq:mm5tl} and sending $\omega$ to infinity, we recover the dS${}_5 \times H^5$ geometry~\eqref{eq:dsto}. In this sense, the $\omega \rightarrow \infty$ limit that leads to 10D Carroll-like geometry at the asymptotic infinity corresponds to the `near-horizon' limit in the bulk, which gets rid of the `1' in the harmonic function $H$.

\vspace{0.5em}\noindent\textbf{Particle Carroll geometry.} The hypothetical duality between instantonic branes and dS space generalizes to other E-branes.
This follows since T-dualizing spatial directions maps ME4T to ME$q$T, which arises from the limit zooming in on a background E$q$-brane in type II${}^*$ string theory~\cite{Blair:2023noj, Gomis:2023eav}.
The resulting Euclidean SYM/matrix theories can then be associated with a holographic description of (conformal) dS${}_{q+1}$. 

In particular, the limit of type IIA* string theory zooming in on a space-filling E9-brane leads to ME9T with a Carrollian target space geometry in the particle sense, \emph{i.e.}~with Carroll boost $\delta_\text{C} x^0 = \beta_i \, x^i$ and $\delta_\text{C} x^i = 0$~\cite{Blair:2023noj}. Now, the prescription~\eqref{eq:mm5tl} is dualised to be~\cite{Blair:2023noj}
\begin{align} \label{eq:mm10tl}
\begin{split}
	\dd s^2 &= - \tfrac{1}{\omega} \, \dd x_0^2 + \omega \, \dd x^i \, \dd x^i, 
        \quad%
    i = 1,\cdots, 9, \\[4pt]
     C^{(9)} &= \tfrac{\omega^2}{g^{}_\text{s}} \, \dd x^1 \cdots \dd x^9\,,
		\qquad\quad\,\,\,\,%
	e^\Phi = \omega^{\frac{5}{2}} \, g^{}_\text{s}\,,
\end{split}
\end{align}
where $C^{(9)}$, rather than $C^{(4)}$, becomes critical. Turning on a finite 9-form fluctuation $c^{(9)}$ around this critical background in ME9T, \emph{s.t.}~the reparametrization of $C^{(9)}$ in Eq..~\eqref{eq:mm10tl} is extended to be 
\begin{align}
    C^{(9)}= \tfrac{\omega^2}{g^{}_\text{s}} \, \dd x^1 \cdots \dd x^9 + c^{(9)}. 
\end{align}
Here, the Hodge dual of $c^{(9)}$ is a concrete string theory realization of the contravariant vector field $M^\mu$ in Carroll geometry~\cite{Hartong:2015xda}. As a consequence we are led to a natural algebraic origin of this extra field, which thus corresponds to the central charge of the space-filling brane in the contracted type IIA${}^*$ algebra. 

\vspace{0.5em}\noindent\textbf{Flat space dual of IKKT matrix theory.}
A particularly interesting case is when we wrap the background E4-brane over a 4-torus and T-dualize in all the brane directions. Applying the standard Buscher rule~\cite{Buscher:1987sk, Buscher:1987qj}, the limiting prescription~\eqref{eq:mm5tl} now gives rise to the following reparametrization of the background fields in type IIB${}^*$ string theory~\cite{Blair:2023noj, Gomis:2023eav}:
\begin{align} \label{eq:mmotl}
    \dd s^2 = \tfrac{1}{\omega} \, \dd x^\mu \, \dd x^{}_\mu\,,
        \quad%
    C^{(0)} = \omega^2 / g^{}_\text{s}\,,
        \quad%
    e^{\Phi} = g^{}_\text{s} / \omega^2\,.
\end{align}
The $\omega \rightarrow \infty$ limit zooms in on a background instanton localized in spacetime and leads to ME0T, which has a Lorentzian target space. 
This can be seen for instance by deriving the form of the F1-string action by T-dualising~\eqref{eq:fsa} as in~\cite{Gomis:2023eav}, obtaining
\begin{align} \label{eq:fsammot}
    S = \frac{T}{2} \int \dd^2 \sigma \, \partial_\tau X^\mu \, \partial_\tau X_\mu\,,
\end{align}
which is Lorentz invariant in the target space. 
This worldsheet theory coincides with the tensionless limit of the relativistic string, if $T$ in Eq.~\eqref{eq:fsammot} is viewed as an effective tension kept fixed when sending $T_{\text{original}}/T$ to zero~\cite{Isberg:1993av}. 
The chiral sector of this tensionless string is related to ambitwistor string theory~\cite{Mason:2013sva, Casali:2016atr, Bagchi:2020fpr}, where string amplitudes on nodal Riemann spheres describe particle scatterings in the CHY formalism~\cite{Geyer:2015bja, Geyer:2018xwu}. 

Nevertheless, just like in BFSS matrix theory, where the fundamental degrees of freedom are the D0-branes, the tensionless string in ME0T does not play any fundamental role, either. Instead, the dynamics of ME0T is captured by the D-instantons (or E0-branes), which are described by IKKT matrix theory~\cite{Ishibashi:1996xs} 
\footnote{This connection between IKKT matrix theory and the tensionless string suggests a string field theoretic description underlying the ME0T dynamics~\cite{Gomis:2023eav}.}
\begin{align}
	S_\text{IKKT} = - \frac{T_0}{4} \, \text{tr}\Bigl( [X^\mu\,,\,X^\nu] \, [X_\mu,\,X_\nu] \Bigr) + \text{fermions}.
\end{align}
This theory is related to the Euclidean YM in Eq.~\eqref{eq:efba} via spacelike T-duality transformations. Holographically, the gravity dual is associated with the following geometry,
\begin{align} \label{eq:sf}
    \dd s^2 = \tfrac{1}{\mathbb{\Omega}} \, \dd x^\mu \, \dd x^{}_\mu\,,
        \quad%
    C^{(0)} = \mathbb{\Omega}^2 / g^{}_\text{s}\,,
        \quad%
    e^{\Phi} = g^{}_\text{s} / \mathbb{\Omega}^2\,,
\end{align}
with $\mathbb{\Omega} = (x^\mu \, x_\mu / \ell_0^2)^2$ and $\ell_0^8 \sim N \, g^{}_\text{s}$\,. This geometry arises from applying the ME0T limiting prescription~\eqref{eq:mmotl}, which we implement equivalently by rescaling $\hat{x}^\mu = x^\mu/\sqrt{\omega}$ and $\hat{g}^{}_\text{s} = g^{}_\text{s} / \omega^2$, to the E0-brane geometry~\footnote{This is the Lorentzian signature version of the D-instanton geometry~\cite{Gibbons:1995vg}.}
\begin{align} \label{eq:ezbg}
\begin{split}
    \dd \hat{s}^2 &= \hat{H}^\frac{1}{2} \, \dd \hat{x}^\mu \, \dd \hat{x}^{}_\mu\,, 
        \qquad%
    \hat{H} = 1 + \hat{\ell}_0^8 \, / \, (\hat{x}^\mu \, \hat{x}_\mu)^4, \\[4pt]
    \hat{C}^{(0)} &= \hat{g}^{-1}_\text{s} \, \hat{H}^{-1},
        \,\,\quad\qquad%
    \hat{e}^\Phi = \hat{g}^{}_\text{s} \, \hat{H},
        \quad%
    \hat{\ell}_0^8 \sim N \, \hat{g}^{}_\text{s}\,,
\end{split}
\end{align}
which, after smearing $\hat{x}^1,\cdots,\hat{x}^4$, is related to the E4-brane geometry in Eq.~\eqref{eq:efbg} via spacelike T-duality. Note that the dual bulk geometry~\eqref{eq:sf} is written in the string frame. In Einstein frame, the bulk geometry is Minkowski space. This duality between IKKT matrix theory and Lorentzian gravity on bulk Minkowski space can be seen as an example of flat space holography.

\vspace{0.5em}\noindent\textbf{Holography with a Carroll-like bulk.}
We have seen that the bulk dS space arises from performing a decoupling limit zooming in on certain background instantonic brane at the asymptotic infinity. It is also possible to perform consecutive decoupling limits asymptotically~\cite{Blair:2024aqz}, which can lead to a bulk Carroll geometry. As an example, we perform an asymptotic ME0T limit in the bulk dS${}_5 \times H^5$ geometry~\eqref{eq:dsto}. 
For this purpose, we first generalize the limiting prescription~\eqref{eq:mmotl} to apply to a curved background.
We introduce ME0T background fields -- the metric $g^{}_{\mu\nu}$\,, $B$-field $b^{(2)}$, RR potentials $c^{(q)}$, and dilaton $\varphi$, and write the defining ME0T limit of a IIB${}^*$ background as
\begin{subequations} \label{eq:cmeztl}
\begin{align}
    \dd s^2 &= \tfrac{1}{\tilde{\omega}} \, g^{}_{\mu\nu} \, \dd x^\mu \, \dd x^\nu \,,
        &%
    C^{(0)} &= \tilde{\omega}^2 \, e^{-\varphi} + c^{(0)}, \\[4pt]
    e^\Phi &= \tfrac{1}{\tilde{\omega}^2} \, e^{\varphi}, 
        \,\,\,\,
    B^{(2)} = b^{(2)},
        &%
    C^{(q)} &= c^{(q)}, \,\,\,\, q \neq 0\,.
\end{align}
\end{subequations}
We have introduced $\tilde{\omega}$ to distinguish from the parameter $\omega$ associated with the original ME4T limit. 

We now apply an asymptotic ME0T limit to the bulk dS${}_5 \times H^5$ geometry in Eqs.~\eqref{eq:dsto}~\&~\eqref{eq:mmftrrgs} by using the rescaling above Eq.~\eqref{eq:ezbg} (with now $\tilde{\omega}$), along with the critical RR potential $C^{(0)} = \tilde{\omega}^2 / g^{}_\text{s}$, which is constant and thus does not affect the validity of this background as a solution of IIB${}^*$ supergravity. This $\tilde{\omega}$  reparametrization of the dS${}_5 \times H^5$ geometry assumes the form of Eq.~\eqref{eq:cmeztl}. Taking $\tilde{\omega} \rightarrow \infty$, the bulk geometry remains Lorentzian. This is the asymptotic ME4T-ME0T limit of the bulk E4-brane geometry~\eqref{eq:efbg}. 

Next, we consider an asymptotic ME5T-ME1T limit of a bulk E5-brane geometry in IIA${}^*$ theory: the asymptotic ME5T limit is aligned with the E5-brane and corresponds to the bulk near-horizon limit that leads to a conformally dS${}_6$ space, while the ME1T limit requires an extra critical RR 1-form that turns out to be pure gauge. The resulting bulk geometry is putatively a solution to the ME1T limit of supergravity, which develops a non-Lorentzian structure. Keeping the ME1T parameter $\tilde{\omega}$, the bulk metric after performing the asymptotic ME5T limit is
\begin{align} \label{eq:meotcg}
    \dd s^2 &= \frac{-\dd x_0^2 + \dd x_6^2 + \cdots + \dd x_9^2}{\tilde{\omega} \, \mathbb{\Omega}} + \frac{\mathbb{\Omega}}{\tilde{\omega}} \bigl( \dd x_2^2 + \cdots + \dd x_5^2 \bigr) \notag \\[4pt]
    &\quad + \tilde{\omega} \, \mathbb{\Omega} \, \dd x_1^2\,,
        \qquad\qquad\qquad\quad\,%
    \mathbb{\Omega} = (\tau / \ell)^\frac{3}{2},
\end{align}
with $\tau^2 = x_0^2 - x_6^2 - \cdots - x_9^2 > 0$\,. Before setting $\tilde{\omega} \rightarrow \infty$, the bulk geometry contains a conformally dS${}_6$ space. In contrast, the 10D metric description is \emph{not} valid anymore after taking $\tilde{\omega}$ to infinity. Instead, the spatial direction $x^1$ becomes absolute, and the 10D geometry is Carroll-like. This example with a bulk Carroll-like geometry generalizes easily to other E-branes, mirroring~\cite{Lambert:2024uue, Lambert:2024yjk, Blair:2024aqz, Lambert:2024ncn, Harmark:2025ikv}. 

The Carroll-like geometry that arises from the $\tilde{\omega} \rightarrow \infty$ of Eq.~\eqref{eq:meotcg} can be thought of as backreacting E5-branes in ME1T. Further backreacting the E1-branes deform this ME1T geometry to be Lorentzian, in a way that is dual to the $T\bar{T}$-deformation~\cite{Blair:2024aqz}. The line element describing this resulting geometry is in form the same as Eq.~\eqref{eq:meotcg}, but now with $\tilde{\omega}$ replaced with $(\tau / \ell' )^{3/2}$. 

\vspace{0.5em}\noindent\textbf{Infinite boosts in M-theory.}
We have shown that Carroll geometry is naturally associated with hypothetical dS holography in string theory.
We now argue that it also appears in the well-studied AdS/CFT correspondence. As an example, we consider the AdS${}_5 \times S^5$ geometry, 
\begin{align} \label{eq:adsto}
\begin{split}
	\dd s^2 &= \mathbb{\Omega} \, \bigl( - \dd x_0^2 + \cdots + \dd x_3^2 \bigr) + \frac{\dd r^2 + r^2 \, \dd \Omega^2_5}{\mathbb{\Omega}}\,, \\[4pt]
    C^{(4)} &= \mathbb{\Omega}^2 \, g^{-1}_\text{s} \, \dd x^0 \cdots \dd x^3,
        \qquad%
    e^\Phi = g^{}_\text{s}\,,
\end{split}
\end{align}
where $\mathbb{\Omega} = r^2 / \ell^2$ and $r^2 = x_4^2 + \dots + x_9^2$\,, with $\ell \sim (N \, g^{}_\text{s})^{\frac{1}{4}}$ the AdS scale. In the $\mathbb{\Omega} \rightarrow 0$ limit, 
we approach the center of the bulk, \emph{i.e.}~the location of the branes before the near-horizon limit was performed. In this limit, the 10D geometry is Carroll-like, which we again probe by the F1-string. The bosonic sector of the Polyakov string is in form the same as Eq.~\eqref{eq:fsa} with a Carroll-like target space geometry, except that now $a = 0, \cdots, 3$ and $i = 4, \cdots, 9$. This is reminiscent of the effective Carrollian structure on black hole and cosmological horizons~\cite{deBoer:2021jej, Bagchi:2023cfp, Bagchi:2024rje}. Note that the RR potential in the $\mathbb{\Omega} \rightarrow 0$ limit is \emph{not} critical, \emph{i.e.}~the resulting corner of string theory does not have a BPS nature compared to the opposite limit where $\mathbb{\Omega} \rightarrow \infty$. In the latter, we zoom in on a background D3-brane by fine tuning its RR charge to cancel the blown-up brane tension. This BPS decoupling limit leads to \emph{Matrix 3-brane Theory} (M3T) whose fundamental degrees of freedom are the D3-branes, underlying the correspondence between $\mathcal{N} = 4$ SYM and gravity on AdS${}_5$~\cite{Blair:2024aqz}. Unlike the Carrollian structure at the center of the bulk, the M3T geometry is Galilei-like, with an absolute Minkowski subspace. 

Analogously, for a dS bulk as in Eq.~\eqref{eq:dsto}, the asymptotic geometry is Carroll-like while the geometry at $\mathbb{\Omega} \rightarrow 0$ is Galilei-like. 
The complementary structures of the AdS and dS cases are depicted visually in Fig.~\ref{fig:rm}. Note that the constant parameter $\omega$ and the background-dependent function $\mathbb{\Omega}$ are exchangeable when the limits of $\omega \rightarrow 0$ and $\omega \rightarrow \infty$ are concerned, due to the emergent dilatational symmetry (see the End Matter)~\cite{Blair:2024aqz}.

The M-theory uplift sheds considerable light on the role of $\omega$. 
Consider the decoupling limit parametrized by the same $\omega$ that leads to BFSS matrix theory on D0-branes in \emph{Matrix 0-brane Theory} (M0T)~\cite{Blair:2023noj, Gomis:2023eav, Blair:2024aqz}, which is T-dual to all cases discussed in the Letter.
Uplifting to M-theory, $\omega \rightarrow \infty$ is
an infinite boost in a compact spatial dimension, leading to a null compactification. \emph{i.e.}~M-theory in the Discrete Light Cone Quantization (DLCQ)~\cite{Susskind:1997cw, Sen:1997we, Seiberg:1997ad}. Since the compactification breaks 11D Lorentz boost symmetry in M-theory, the infinite boost should be regarded as a decoupling limit rather than a mere change of frame.

We now show how $\omega$ is related to the boost parameter in M-theory. Compactify $x^{10}$ over a circle of radius $R_\text{s}$. Boosting in $x^{10}$ with rapidity $\textbf{w}$ gives
\begin{align}
    \begin{pmatrix}
        \tilde{x}^0 \\[4pt]
        \tilde{x}^{10}
    \end{pmatrix}
    =
    \begin{pmatrix}
        \cosh \textbf{w} &\,\, -\sinh \textbf{w} \\[4pt]
        - \sinh \textbf{w} &\,\, \cosh \textbf{w}
    \end{pmatrix}
    \begin{pmatrix}
        x^0 \\[4pt]
        x^{10}
    \end{pmatrix}.
\end{align}
The boost velocity is $v = \tanh \textbf{w}$. 
Define~\cite{Elizalde:1976vj, Blair:2024aqz},
\begin{align}
    x^- \! = \tfrac{1}{\sqrt{2}} \bigl( \tilde{x}^{10} \! - \tilde{x}^0 \bigr)\,,
        \quad%
    x^+ \! = \tfrac{1}{\sqrt{2}} \bigl( \tilde{x}^0 \! + \tilde{x}^{10} \bigr) - \tfrac{x^-}{2 \, \omega^2}\,,
\end{align}
with $\omega = e^\textbf{w} / \sqrt{2}$. These `almost' lightlike coordinates are constructed such that only $x^-$ is periodic, with radius $R = \omega \, R_\text{s}$. In the $\omega \rightarrow \infty$ limit, we are led to DLCQ M-theory by simultaneously sending $R_\text{s}$ to zero, such that the effective radius $R$ of the lightlike circle in $x^-$ is kept finite. Dimensionally reducing DLCQ M-theory along $x^-$ to type IIA superstring theory gives rise to the limiting prescription defining M0T at $\omega \rightarrow \infty$~\cite{Gomis:2023eav, Blair:2023noj, Blair:2024aqz}:
\begin{align} \label{eq:mztl}
\begin{split}
    \dd s^2 &= - \omega \, \dd x_0^2 + \omega^{-1} \bigl( \dd x_1^2 + \cdots \dd x_9^2 \bigr)\,, \\[4pt]
    C^{(1)} &= \omega^2 \, g^{-1}_\text{s} \, \dd x^0,
        \qquad%
    e^\Phi = \omega^{- \frac{3}{2}} g^{}_\text{s}\,.
\end{split}
\end{align}
We have seen that DLCQ M-theory arises from the limit under which the rapidity \textbf{w} is sent to positive infinity. As expected, a timelike T-duality transformation of Eq.~\eqref{eq:mztl} leads to the ME0T prescription~\eqref{eq:mmotl}, which underlies the timelike T-dual between BFSS and IKKT matrix theory. 

There is yet another infinite boost in which $\textbf{w} \rightarrow - \infty$\,, \emph{i.e.}~$\omega \rightarrow 0$. Requiring that the $x^-$ circle be finite, we have to set $R_\text{s} \rightarrow \infty$\,, which effectively decompactifies the original circle in $x^{10}$. This $\omega \rightarrow 0$ limit leads, after applying duality transformations, to a web of Carroll-like theories in 11D and 10D~\footnote{See~\cite{Chen:2023pqf, Majumdar:2024rxg, Bagchi:2024epw} for different constructions of Carroll physics by focusing on the light cone in field theories.}. This new duality web can be used to guide the classification of Carrollian field theories~\cite{carrolldual}, which goes beyond standard effective field theory and still lacks an overarching  principle. Understanding these may provide a new angle, complementary to BFSS matrix theory, on approaching M-theory in asymptotically flat spacetime. However, since the BPS properties do not pertain in this other infinite boost limit, it may be difficult to gain much non-perturbative control over such Carroll-like corners. Moreover, it is also curious to point out that, in the $\omega \rightarrow 0$ limit associated with a background D9-brane, the probe F1-string also coincides with the tensionless string described by Eq.~\eqref{eq:fsammot}. 

It also follows that ME0T arises from compactifying M-theory over an exotic 2-torus with both cycles lightlike. The associated SL($2,\mathbb{Z}$) transformations exhibit duality asymmetry~\cite{Bergshoeff:2022iss, Bergshoeff:2023ogz, Ebert:2023hba, Blair:2025prd}, suggesting a relation between the $\omega \rightarrow \infty$ and $\omega \rightarrow 0$ limits. This would be interesting to study further in the context of timelike dualities and negative branes~\cite{Hull:1998vg, Hull:1998fh, Hull:1999mt, Dijkgraaf:2016lym, Berman:2019izh}. 

\vspace{0.5em}\noindent\textbf{Outlook.}
In this Letter, we re-examined the role of string theory in the holographic constructions of dS and flat space.
This built on Hull's insights from type II${}^*$ string theory via timelike T-duality~\cite{Hull:1998vg}, and the recently developed non-Lorentzian perspective on AdS/CFT~\cite{Blair:2024aqz}. 
In parallel with non-Lorentzian supergravity (see \emph{e.g.}~\cite{Bergshoeff:2019pij, Bergshoeff:2021tfn, Bergshoeff:2024nin}), it would be interesting to find the bulk Carroll-like theories that arise in the limits considered in this Letter, leading to stringy generalisations of  
 Carroll gravity actions~\cite{Henneaux:1979vn, Hartong:2015xda, Bergshoeff:2017btm, Gallegos:2020egk, Hansen:2021fxi, Figueroa-OFarrill:2022mcy}.
We have focused on dS in 10D string theory, but M-theory examples can also be explored, combining~\cite{Hull:1998vg, Hull:2001ii} and~\cite{Blair:2023noj, Blair:2024aqz}.
It was further shown in~\cite{Blair:2024aqz} that undoing BPS decoupling limits defines higher-dimensional generalisations of the $T\bar{T}$ deformation, which would lead here to a geometric flow equation in dS space. 

The type II${}^*$ supergravity and the instantonic field theory duals exhibit various instabilities, which might indicate the possibility of a vacuum decay (\emph{e.g.}~to AdS space), if type II${}^*$ theories could also be considered as vacua in M-theory. From this perspective, the quest for a stable dS vacuum in the string landscape may be misleading, and the holographic description of flat and dS space in terms of instantons could imply the start of vacuum decay~\footnote{The decay probability tending to unity asymptotically in such a metastable configuration can be viewed as an argument against any CFT dual. Indeed, the alleged dual instantonic branes signal the initiation of vacuum decay~\cite{Danielsson:2018ztv}.}.  

Furthermore, the connection between our framework and tensionless string theory may also underlie the dS/CFT proposals using Vasiliev's higher-spin gravity~\cite{Anninos:2011ui, Anninos:2017eib}, following the insights from~\cite{Sundborg:2000wp} for AdS/CFT. 

Taking seriously the dual Euclidean SYM corresponding to the dynamics on the bulk dS background may provide an alternative way of computing correlation functions in the late Universe~\footnote{See~\emph{e.g.}~\cite{McFadden:2009fg, McFadden:2010na, Afshordi:2016dvb} for using holography to predict CMB observables.}. It is also valuable to explore further the well-defined sector from topologically twisting type II$^*$ strings~\cite{Hull:1998vg} and instanton contributions analogous to~\cite{Witten:2003nn} and related self-dual reincarnation~\cite{Kmec:2025ftx}.
This may in turn benefit the program of flat holography, as dS space arises from slicing Minkowski space~\cite{deBoer:2003vf}. Another interesting perspective is that IKKT matrix theory may capture the dynamics of a putative 9D Carrollian field theory dual of the 10D flat space. Evidence for this comes from BFSS matrix theory, which can be obtained from a Carrollian limit of 10D $\mathcal{N}=1$ SYM~\cite{rsgbfss}.

We hope the new relations studied in this Letter could help reevaluate what string theory may offer for many of the important puzzles associated with de Sitter space and Carrollian physics. 

\vspace{3mm}

\noindent 
{\bf Acknowledgements:} We thank Joaquim Gomis, Emil Have, Chris Hull, Johannes Lahnsteiner, and Florian Niedermann for useful discussions. Z.Y. would like to thank the participants from the workshop on \emph{From Asymptotic Symmetries to Flat Holography: Theoretical Aspects and Observable Consequences} at the Galileo Galilei Institute, Florence (June, 2025) for stimulating interactions.
The work of C.B. is supported through the grants CEX2020-001007-S and PID2021-123017NB-I00, funded by MCIN/AEI/10.13039/501100011033 and by ERDF A way of making Europe.
The work of N.O. is supported in part by VR project Grant 2021-04013 and Villum Foundation Experiment Project No.~00050317. 
The work of Z.Y. is
supported in part by Olle Engkvists Stiftelse project Grant 234-0342 and the Villum Young Investigator Programme under project No.~71589. The Center of Gravity is a Center of Excellence funded by the Danish National Research Foundation under grant No.~184. 

\bibliography{cgmds}

\onecolumngrid

\clearpage

\vspace{0.4cm}
\noindent
\begin{center}
    \large \textbf{End Matter}
\end{center}
\label{EndMatter} 

\quad

\twocolumngrid

\noindent 
Here we discuss the action for a probe F1-string in the dS${}_5 \times H^5$ geometry~\eqref{eq:dsto}, in order to show how non-Lorentzian symmetries manifest when we take the string to the asymptotic infinity and center of the bulk, respectively. This requires the framework developed in~\cite{Gomis:2023eav}, which we first briefly review below before describing the probe string in dS space. 

\vspace{0.5em}\noindent\textbf{Strings for BFSS matrix theory.} We first consider the F1-string in \emph{Matrix 0-brane Theory} (M0T), where the D0-brane dynamics is described by BFSS matrix theory. The M0T limiting procedure is given by Eq.~\eqref{eq:mztl}, where $\omega$ is sent to infinity. 
This can be viewed as a non-relativistic limit where the speed of light is sent to infinity.
To see this, consider restoring the dependence on the speed of light $c$ in the conventional Nambu-Goto string action:
\begin{align} \label{eq:ngnrl}
\begin{split}
    S^{}_\text{NG} &= - \frac{T}{c} \int \dd^2 \sigma \, \sqrt{-\det G_{\alpha\beta}}\,, \\[4pt]
    G^{}_{\alpha\beta} &\equiv - c^2 \, \partial^{}_\alpha X^0 \, \partial^{}_\beta X^0
        + \partial^{}_\alpha X^i \, \partial^{}_\beta X^i. 
\end{split}
\end{align}
Note that the worldsheet time $\tau$ also acquires a prefactor $c$, \emph{i.e.}~$\sigma^\alpha = (c \, \tau, \sigma)$. However, this dependence on $c$ precisely cancels out in the Nambu-Goto action. It can be shown that Eq.~\eqref{eq:ngnrl} is equivalent to the Nambu-Goto action in the background~\eqref{eq:mztl} defining the M0T limit, upon identifying $c$ (in natural units) with $\omega$. Therefore, the M0T limit is an infinite $c$ limit. 
A short calculation shows that in this limit the string action is finite, 
\begin{align} \label{eq:ngmzt}
    S^\text{M0T}_\text{NG} = - T \!\int \dd^2 \sigma \, \sqrt{-\det \!
    \begin{pmatrix}
        0 
        &\,\, \partial^{}_\beta X^0 \\[4pt]
        \partial^{}_\alpha X^0 &\,\, \partial^{}_\alpha X^i \, \partial^{}_\beta X^i
    \end{pmatrix}}\,,
\end{align}
in terms of a three-by-three determinant. 
Next, we consider the Polyakov formulation for the M0T string. We start with the standard Polyakov action,
\begin{align}
    S^{}_\text{P} = - \frac{T}{2 \, c} \int \dd^2 \sigma \, \sqrt{-\det h} \, h^{\alpha\beta} \, G^{}_{\alpha\beta}\,, 
\end{align}
with $G_{\alpha\beta}$ given in Eq.~\eqref{eq:ngnrl}. 
Keeping explicitly the dependence on the speed of light $c$, we introduced the worldsheet metric $h_{\alpha\beta} = - c^2 \, e_\alpha{}^0 \, e_\beta{}^0 + e_\alpha{}^1 \, e_\beta{}^1$, with $h \equiv \det h_{\alpha\beta}$ and $h^{\alpha\beta}$ the inverse metric. It follows that
\begin{align}
\begin{split}
    S^{}_\text{P} = \frac{T}{2} \int & \dd^2 \sigma \, e \, \Bigl[ c^2 \bigl( e^\alpha{}^{}_1 \, \partial^{}_\alpha X^0 \bigr)^2 - \bigl( e^\alpha{}^{}_0 \, \partial^{}_\alpha X^0 \bigr)^2 \\[4pt] 
    & - \bigl( e^\alpha{}^{}_1 \, e^\beta{}^{}_1 - c^{-2} \, e^\alpha{}_0 \, e^\beta{}_0 \bigr) \, \partial^{}_\alpha X^i \, \partial^{}_\beta X^i \Bigr]\,,
\end{split}
\end{align}
with $e \equiv \det (e_\alpha{}^0 \,\,\, e_\alpha{}^1)$\,. 
We then rewrite this action using a Hubbard-Stratonovich transformation, by integrating in an auxiliary field $\lambda$\,, such that
$c^2 ( e^\alpha{}^{}_1 \, \partial^{}_\mu X^0 )^2$ is replaced with $\lambda \, e^\alpha{}^{}_1 \, \partial^{}_\mu X^0 - \tfrac{1}{4} \, c^{-2} \, \lambda^2$\,.
In the infinite speed of light limit, we are led to the M0T string action,
\begin{align} \label{eq:mztsp}
\begin{split}
    S^\text{M0T}_\text{P} \! = \frac{T}{2} \int & \dd^2 \sigma \, e \, \Bigl[ \lambda \, e^\alpha{}^{}_1 \, \partial^{}_\alpha X^0 - \bigl( e^\alpha{}^{}_0 \, \partial^{}_\alpha X^0 \bigr)^2 \\
    & \qquad\qquad\qquad - e^\alpha{}^{}_1 \, e^\beta{}^{}_1 \, \partial^{}_\alpha X^i \, \partial^{}_\beta X^i \Bigr]\,.
\end{split}
\end{align}
Integrating out the Lagrange multiplier $\lambda$ together with the auxiliary zweibein fields $e_\alpha{}^0$ and $e_\alpha{}^1$, the Nambu-Goto formulation~\eqref{eq:ngmzt} is reproduced. The worldsheet geometry here is Galilean. In imaginary time, the worldsheet topology is captured by the nodal Riemann spheres. If we instead parametrize the worldsheet metric as
\begin{equation} \label{eq:hrc}
    h_{\alpha\beta} = - e_\alpha{}^0 \, e_\beta{}^0 + c^2 \, e_\alpha{}^1 \, e_\beta{}^1, 
\end{equation}
we are led to a Carroll worldsheet after the $c \rightarrow \infty$ limit is performed, which is more commonly used in the literature. This Carroll choice is also what we used in the main part of the paper. For our purposes here, the choice between the Galilei and Carroll parametrization of the worldsheet geometry is merely a convention. We will therefore stick to the Carroll choice in the rest of this appendix, just as in the main part of the Letter.  

The M0T string is also related~\cite{Gomis:2023eav} to the non-vibrating Galilei string~\cite{Batlle:2016iel, Gomis:2016zur} and tropological sigma models for Gromov-Witten invariants~\cite{Albrychiewicz:2023ngk}. 

\vspace{0.5em}\noindent\textbf{Dilatation symmetry.} We now consider the M0T string in curved backgrounds by covariantising as
\begin{align}
    \dd x^0 \rightarrow \dd x^\mu \, \tau^{}_\mu\,,
        \qquad%
    \dd x^i \rightarrow \dd x^\mu \, E^{}_\mu{}^i\,,
\end{align}
where $\tau^{}_\mu$ and $E^{}_\mu{}^i$ are temporal and spatial vielbein fields, respectively. These vielbein fields encode the target space Galilei geometry. The M0T string~\eqref{eq:mztsp} generalizes to
\begin{align} \label{eq:mztspcb}
    S^{}_\text{M0T} = \frac{T}{2} \int & \dd^2 \sigma \, e \, \Bigl[ \lambda \, e^\alpha{}^{}_0 \, \partial^{}_\alpha X^\mu \, \tau^{}_\mu + \bigl( e^\alpha{}^{}_1 \, \partial^{}_\alpha X^\mu \, \tau^{}_\mu \bigr)^2 \notag \\[4pt] 
    & \quad + e^\alpha{}^{}_0 \, e^\beta{}^{}_0 \, \partial^{}_\alpha X^\mu \, \partial^{}_\beta X^\nu \, E^{}_\mu{}^i \, E^{}_\nu{}^i \Bigr]\,.
\end{align}
We have here switched to the Carroll worldsheet by applying the mapping $e^{}_\alpha{}^0 \rightarrow i \, e^{}_\alpha{}^1$, $e^{}_\alpha{}^1 \rightarrow - i \, e^{}_\alpha{}^0$, and $\lambda \rightarrow - i \, \lambda$. 
This action is manifestly invariant under an emergent dilatation symmetry, which acts as
\begin{subequations}
\begin{align}
    \tau^{}_\mu &\rightarrow \Delta^{1/2} (X) \, \tau^{}_\mu\,, 
        &%
    e^{}_\alpha{}^1 &\rightarrow \Delta (X) \, e^{}_\alpha{}^1\,, \\[4pt]
    E^{}_\mu{}^i &\rightarrow \Delta^{-1/2} (X) \, E^{}_\mu{}^i, 
        &%
    \lambda &\rightarrow \Delta^{-3/2} (X) \, \lambda\,.
\end{align}
\end{subequations}
This dilation symmetry implies that the $\omega \rightarrow \infty$ limit is insensitive to whether $\omega$ is spacetime dependent, and justifies promoting $\omega$ to $\mathbb{\Omega} (X)$ in the bulk of this Letter. 

\vspace{0.5em}\noindent\textbf{Probe string in de Sitter space.} We are now ready to study a probe string in the dS${}_5 \times H^5$ geometry~\eqref{eq:dsto}. Now, $\mathbb{\Omega} = (\tau / \ell)^2$ in Eq.~\eqref{eq:dsto} replaces the constant $\omega$ in the ME4T limiting prescription~\eqref{eq:cmeztl}. Since ME4T is related to M0T via (timelike) T-duality~\cite{Blair:2023noj, Gomis:2023eav}, which only acts on the background fields rather than the worldsheet, the worldsheet metric has to be reparametrized as in Eq.~\eqref{eq:hrc}, but now with $c$ being replaced with $\mathbb{\Omega}$\,, \emph{i.e.}
\begin{equation} \label{eq:hmbo}
    h^{}_{\alpha\beta} = -e^{}_\alpha{}^0 \, e^{}_\beta{}^0 + \mathbb{\Omega}^2 \, e^{}_\alpha{}^1 \, e^{}_\beta{}^1\,.
\end{equation}
Plugging Eqs.~\eqref{eq:dsto} and \eqref{eq:hmbo} into the standard Polyakov string action, we obtain
\begin{align} \label{eq:sds}
\begin{split}
    S^{}_\text{P} &= \frac{T}{2} \int \dd^2 \sigma \, e \, \Bigl[ \mathbb{\Omega}^2 \, \bigl( e^\alpha{}_0 \, \partial^{}_\alpha X^i \bigr)^2 - \bigl( e^\alpha{}_1 \, \partial_\alpha X^i \bigr)^2 \\[4pt] 
    & \hspace{1.6cm} + \bigl( e^\alpha{}_0 \, \partial_\alpha X^a \bigr)^2 - \mathbb{\Omega}^{-2} \, \bigl( e^\alpha{}_1 \, \partial_\alpha X^a \bigr)^2 \Bigr]\,,
\end{split}
\end{align}
which we rewrite equivalently by integrating in an auxiliary field $\lambda_i$\,, such that
\begin{align} \label{eq:mmfts}
    S^{}_\text{P} & \!=\! \frac{T}{2} \!\int\! \dd^2 \sigma \, e \, \biggl\{ \! \bigl( e^\alpha{}_0 \, \partial_\alpha X^a \bigr)^{\!2}\!\! - \! \bigl( e^\alpha{}_1 \, \partial_\alpha X^i \bigr)^{\!2}\!\!+ \! \lambda_i \,  e^\alpha{}_0 \, \partial^{}_\alpha X^i \notag \\[4pt] 
    & \hspace{2.1cm} - \mathbb{\Omega}^{-2} \, \Bigl[ \tfrac{1}{4} \, \lambda_i \, \lambda^i + \bigl( e^\alpha{}_1 \, \partial_\alpha X^a \bigr)^2 \Bigr] \bigg\}\,.
\end{align}
Recall that it is the index $a$ that includes the time direction. 
In the asymptotic infinite limit, we set $\mathbb{\Omega} \rightarrow \infty$, which leads to the ME4T string action described by the first line in Eq.~\eqref{eq:mmfts}. This resulting action describes a string in 10D Carroll-like geometry and reduces to Eq.~\eqref{eq:fsa} in conformal gauge. Alternatively, taking the $\mathbb{\Omega} \rightarrow 0$ limit in Eq.~\eqref{eq:sds}, the string is sent towards the center of the bulk geometry.
Performing a Hubbard-Stratonovich transformation by introducing an auxiliary field $\lambda_a$, we find in conformal gauge that
\begin{align} \label{eq:sdsr}
\begin{split}
    S^{}_\text{P} &\rightarrow \frac{T}{2} \int \dd^2 \sigma \, \Bigl[ \bigl( \partial_\tau X^a \bigr)^2 - \bigl( \partial_\sigma X^i \bigr)^2 + \lambda_a \, \partial_\sigma X^a \Bigr].
\end{split}
\end{align}
This action is invariant under the target space Galilei-like boost $\delta_\text{G} X^a = 0$ and $\delta_\text{G} X^i = v^i{}_a \, X^a$. Note that the Lagrange multiplier $\lambda_a$ also transforms accordingly. The worldsheet is now Galilean instead of Carrollian. 

\vspace{0.5em}\noindent\textbf{Strings for IKKT matrix theory.} Finally, we consider a probe string in the conformally flat spacetime~\eqref{eq:sf}, where the field theory dual is supposedly IKKT matrix theory. Following the same arguments, we find that, in conformal gauge, the string at the asymptotic infinity with $\mathbb{\Omega} \rightarrow \infty$ is the tensionless string~\eqref{eq:fsammot}, whereas at the center of the bulk, with $\mathbb{\Omega} \rightarrow 0$, the string action becomes 
\begin{align} \label{eq:mezta}
    S^{}_\text{P} \rightarrow \frac{T}{2} \int \dd^2 \sigma \, \Bigl( \partial_\tau X^\mu \, \partial_\tau X^{}_\mu + \lambda_\mu \, \partial_\sigma X^\mu \Bigr)\,,
\end{align}
with a Lorentzian target space but a Galilean worldsheet. 

It was also shown in~\cite{Gomis:2023eav} that the ME0T string~\eqref{eq:fsammot} (\emph{i.e.}~tensionless string) is related to the ME4T string~\eqref{eq:fsa} via T-duality over a spatial 4-torus and to the M0T string via T-duality over a timelike circle.

We note that the same tensionless string action~\eqref{eq:mezta} also arises from the $\omega \rightarrow 0$ limit of type IIB superstring theory reparametrized as~\cite{Gomis:2023eav, Blair:2023noj}
\begin{align*}
    \dd s^2 \! = \omega \, \dd x^\mu \dd x_\mu, 
        \quad%
    C^{(10)} \! = \tfrac{\omega^2}{g^{}_\text{s}} \, \dd x^0 \!\cdots \dd x^9,
        \quad%
    e^\Phi \! = \omega^{3} g^{}_\text{s}\,. 
\end{align*}
The associated $\omega \rightarrow \infty$ limit zooms in on a background D9-brane and gives rise to \emph{Matrix 9-brane Theory} (M9T). 
However, even though the metric part of the F1-string action coincides, the $\omega \rightarrow \infty$ limit of the above M9T prescription is distinct from ME0T, as the dilaton and RR fields are parametrized differently.

\end{document}